# Digestible Pieces: comparing three options for partitioning the Northeast Pacific Coast for S2S sea surface height prediction

Laura Thapa[a], Marybeth Arcodia[b,c], Elizabeth Barnes[a,d,e]

[a]*Colorado State University, Department of Atmospheric Science, Fort Collins, CO*
[b]*University of Miami Rosenstiel School of Marine, Atmospheric and Earth Science, Miami, FL*
[c]*Frost Institute for Data Science and Computing, Miami, FL*
[d]*Faculty of Computing & Data Sciences, Boston University, Boston, MA*
[e]*Department of Earth & Environment, Boston University, Boston, MA*

*Corresponding author*: Laura Thapa, laura.thapa@colostate.edu

ABSTRACT

We discuss the utility of applying clustering as a pre-processing step for identifying subseasonal-to-seasonal forecasts of opportunity of coastal sea level using convolutional neural networks (CNNs). Clustering leverages potential covariance among points along the same coastline or in the same ocean basin. To evaluate the utility of clustering for reliably identifying forecasts of opportunity, we compare CNNs trained to predict sea level probability distributions in three ways: over the whole Northeast Pacific Coast simultaneously, over pre-determined clusters within this coastline, and at individual gridpoints near tide gauges. All CNN prediction tasks (Whole Coast, Cluster, Point), outperform climatology by a similar margin at Week 3 when the entire test set is used to evaluate CNN skill. However, when comparing the skill of each tasks' 20% most confident predictions, we find the skill of the Cluster and Point tasks to be on par with each other and substantially more skillful than the Whole Coast task. Of the Cluster and Point task, the Cluster task represents all gridpoints in the Northeast Pacific Coast with minimal tunable parameters. Throughout this exercise we learned that clustering gridpoints as a pre-processing step is the preferred approach between the three for making S2S predictions of coastal sea level.

SIGNIFICANCE STATEMENT

When building CNNs to make predictions for geophysical variables, researchers need to decide how to partition the domain of interest. This paper explores three ways to partition the Northeast Pacific Coastline to make subseasonal to seasonal predictions of sea level. We find that pre-clustering the points and training CNNs to predict those clusters yields skillful predictions which can cover the entire domain of interest with a minimal number of CNNs. This work underscores the importance of clustering as a method for dimensionality reduction in geophysical systems and applies that clustering based on spatial or temporal covariability can lead to increased predictive skill.

## 1. Introduction

Machine learning (ML) methods are becoming a common way to study sea level rise and variability over a wide range of temporal and spatial scales (e.g. Ayinde et al., 2023; Bellinghausen et al., 2024; Bruneau et al., 2020; Radin et al., 2024; Sinha et al., 2024;

Tiggeloven et al., 2021). Recent work has shown the importance of partitioning the spatial domain of interest to give models an advantage when making predictions (Radin et al., 2024; Sinha et al., 2024). For instance, when making predictions of non-tidal residual or storm surge at a given tide gauge location (Bellinghausen et al., 2024; Bruneau et al., 2020; Tiggeloven et al., 2021) or predicting the median or mean sea level anomaly over a region (Ayinde et al., 2023; Nieves et al., 2021), it is intuitive to train individual models for each tide gauge separately . For spatial domains containing multiple locations of interest, recent work has shown that covariance among locations along a shared coastline or within a shared ocean basin may depend on the spatial or temporal scale of the available data. For instance, Oelsmann et al., (2024) highlighted the West Coast of North America as a region of coherent variability on a monthly basis and Radin et al., (2024) showed that several sub-basins of the Mediterranean sea co-vary in terms of monthly sea surface temperature. This covariance has been leveraged to improve one month to three year predictions of sea level in the Mediterranean (e.g. Radin et al., 2024) as well as global trends of sea level (e.g. Sinha et al., 2024). In spite of this, recent work in basin- or coastline-scale prediction of sea level is still split on whether to predict locations individually (Bellinghausen et al., 2024), in clusters (Radin et al., 2024) or over all locations in the domain of interest simultaneously (Brettin et al., 2025).

In this work, we explore the lesson we learned for how to partition the Northeast Pacific Coast (NEPC, between 20N and 50N) for making subseasonal-to-seasonal (S2S) predictions of sea level using convolutional neural networks (CNN) trained on climate model data. We define three prediction tasks: Whole Coast, Cluster, and Point. For making Week 3 predictions of sea surface height anomalies, we find that pre-clustering (Cluster task) the points is our preferred choice as compared with predicting the whole NEPC simultaneously (Whole Coast task) or predicting model gridpoints individually (Point task). Clustering produces skillful predictions for all climate model gridpoints along the NEPC with minimal tunable parameters, while the Whole Coast task produces less skillful forecasts and the Point task requires over two orders of magnitude more tunable parameters. Finally, the Cluster task reliably identifies S2S forecasts of opportunity (FOO) for sea level, also known as states of the climate where this variable is more predictable (Mariotti et al., 2020).

## 2. Data and Methods

*a. CESM-HR PI-Control, daily data*

This work uses a set of Community Earth System Model simulations run with a 0.25-degree atmosphere and a 0.1-degree ocean and climate forcing kept at preindustrial (pre-1850) levels for the duration of the simulation period (CESM-HR, PI-Control, Chang et al., 2020). Due to model spinup and data storage constraints, our study uses a total of 52 years of PI-Control daily data. The first 36 years are used for model training, the next 8 years for model validation, and the final 8 years for model testing, equating to a roughly 70%/15%/15% train/validation/test split.

From the CESM-HR daily data, we chose several variables for use in pre-clustering and in the CNNs. The variables are as follows: sea surface temperature (SST), sea level pressure (PSL), precipitation (PRECT), zonal and meridional wind stress (TAUX, TAUY), and sea sea surface height (ETA). An inverse barometer correction was applied to the CESM-HR sea surface height values to account for how pressure anomalies at sea level drive sea surface height anomalies (Brettin et al., 2025). For the pre-clustering, 0.1 degree coastal ETA data was put through a 15-degree, 100-day high pass Butterworth filter in order to make sure clustering was done based on S2S scale covariance. For each CNN input variable, anomalies were computed by removing a smoothed seasonal cycle at each gridpoint, where the smoothed seasonal cycle is defined as the 28-day running mean of the 365 day climatology generated from all 52 years of daily data. Prior to input into the CNNs, the CESM-HR data were regridded to 1-degree for ETA and 2.5-degrees for all other variables and scaled by constants to set the range of each variable roughly between -1 and 1 while preserving spatial gradients. For the CNN target variable, ETA at 0.1 degree resolution along the NEPC (between 20N and 50N) were selected and smoothed for using a 7-day forward mean to compute the Week 3 average inverse barometer corrected sea surface height (note no highpass filtering was applied).

*b. CNN Architecture*

In this work we train convolutional neural networks (CNNs) to ingest global maps of ETA, SST, PSL, PRECT, TAUX, and TAUY to predict probability distributions of ETA anomalies during Week 3 along the NEPC. The global inputs are fed into the model in two channels. The high resolution (1-degree) channel contains ETA, and the low resolution (2.5-degree) channel contains all other variables (Figure 1a). The high and low resolution channels go through padding, convolution, and max pooling in parallel (Figure 1b, pink, green, yellow boxes), and the output of the two parallel processes are concatenated together (Fig 1b, dark blue). Finally the concatenated layer is fed into three sets of dense layers with dropout, each of which predicts one of three parameters (Figure 1b, cyan and lavender) of a sinh-arcsinh-normal (SHASH) distribution (Barnes et al., 2023). The predicted SHASH curves (Figure 1c) may have variable skew ($\gamma$), position ($\mu$), and spread ($\sigma$). During training, the CNNs optimize a negative log-likelihood loss function, allowing the envelope of the predicted SHASH to adjust to a reasonable shape for a given target value.

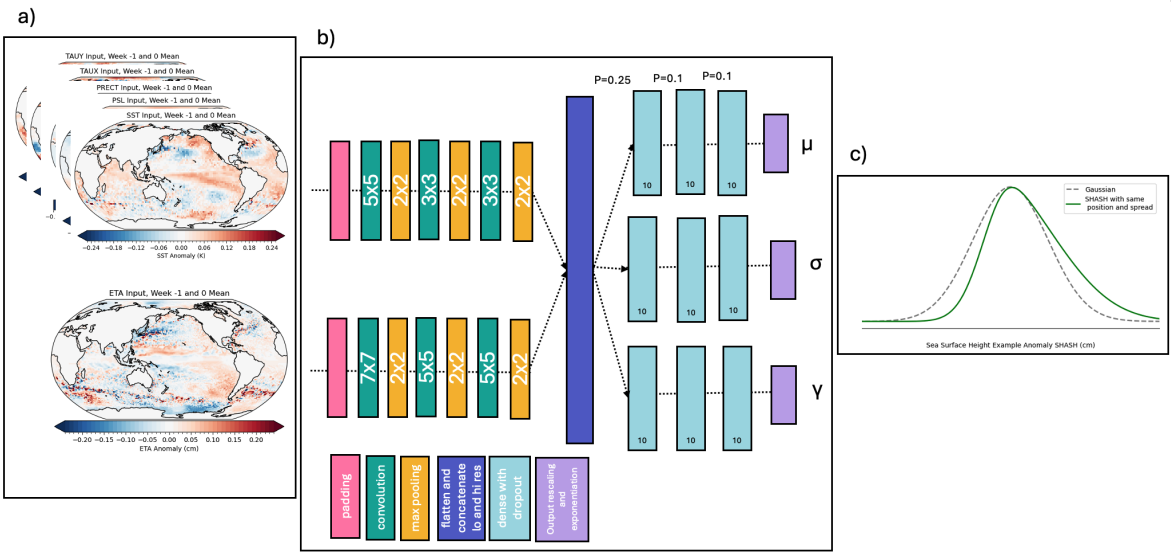

Figure 1. Illustration of model workflow and architecture. a) The two channels of input data. Normalized anomalies of SST, PSL, PRECT, TAUX, and TAUY are provided at low (2.5-degree) resolution, and normalized anomalies of ETA are provided at high (1-degree) resolution. b) CNN architecture. The low- and high-resolution channels are passed through padding (pink) and then through three couplets of convolution and pooling (green and yellow), with the kernel sizes noted in the numbers. The output of the two parallel convolution streams are concatenated together (dark blue), and the concatenated layer is given as input to three sets of dense layers with dropout (light blue, P is the chance an artificial neuron may be dropped), which output the three SHASH parameters (lavender). c) An example SHASH prediction (green line) and the Gaussian with the same spread and position for comparison (dotted gray).

*c. Three prediction tasks*

Our target variable is 0.1-degree ETA values along the NEPC from CESM-HR, which amounts to needing to predict for 650 gridpoints. To investigate the impact of clustering on prediction skill and forecast of opportunity identification, we define three different prediction tasks (Figure 2). In each approach, CNN inputs and hyperparameters do not change, but the way we group outputs does change. In the first prediction task (Whole Coast, Figure 2a) a single CNN is optimized to predict all 650 gridpoints of the NEPC at once. In the second prediction task (Cluster, Figure 2b), fifteen CNNs are trained to predict fifteen separate sections of the NEPC. These separate sections were identified *a-priori* by applying a clustering algorithm developed by Falasca et al., (2024) to the daily, 100-day highpassed ETA data along the NEPC. This clustering algorithm creates communities of coastal points based on ETA covariance and physical distance between the points. Figure 2b shows examples of the clusters determined using the Falasca et al., (2024) method, for example several broad clusters (green, purple, salmon) located north of San Francisco. We evaluate the Whole Coast and Cluster tasks at 35 gridpoints along the coast, corresponding to the locations of tide gauges (Figure 2, black dots). Additionally, individual CNNs are trained to make ETA predictions at each of 35 tide gauge-adjacent gridpoints along the coast (Point, Figure 2c) to have a point of comparison for how a model performs when trained to predict for a single tide gauge gridpoint.

We evaluate the skill and ability to identify forecasts of opportunity (FOO) of the Whole Coast, Cluster, and Point tasks used to partition the outputs. We use continuous ranked probability score (CRPS; Hersbach, 2000), a probabilistic extension of the mean absolute error metric that evaluates reasonableness of a predicted distribution, to evaluate model skill. To identify FOO, we leverage the fact that the CNNs make probabilistic predictions with associated interquartile ranges (IQR). We examine CRPS for all predicted SHASH distributions and for the most confident predictions, i.e. the subset of all predicted SHASH distributions with the 20% narrowest interquartile range. The 20% most confident predictions for each evaluation gridpoint represent our FOOs.

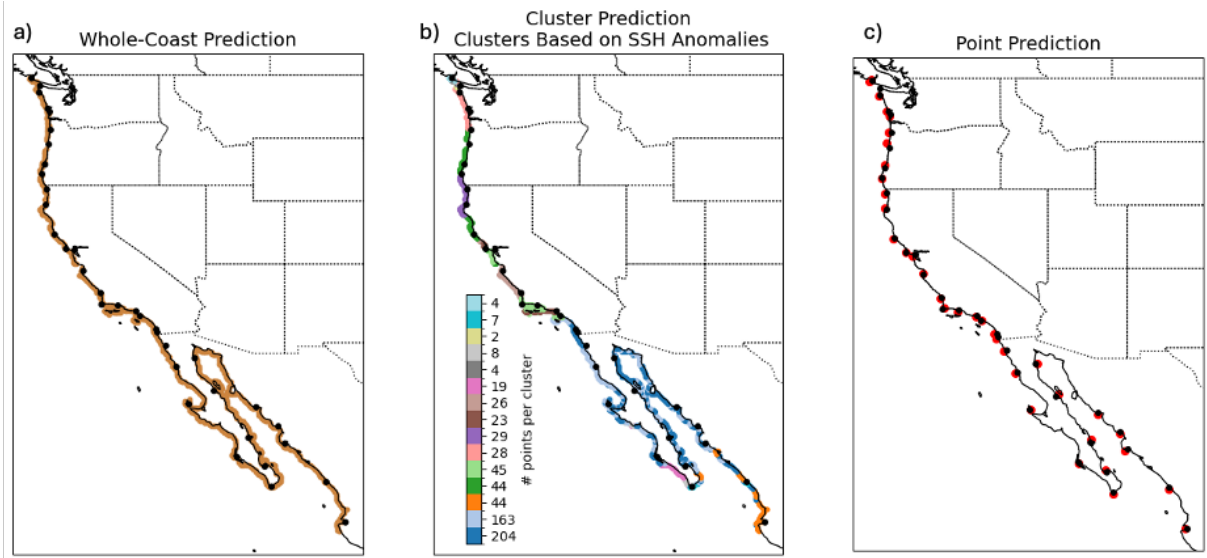

Figure 2. Three Prediction Tasks. In this study, we train the NNs to predict the coastal Week 3 SSH in 3 different ways. For a) and b) gridpoints that are the same color are predicted simultaneously by the same NNs. Therefore, the setup in a) is one single model optimized to predict the whole coast (Whole Coast), and the setup in b) is training 15 models to predict clusters of gridpoints which are the same color (Cluster). In c) an individual NN is trained to make predictions at each red gridpoints (Point). Black dots ion a)-c) represent the location of tide gauges, where the model will be evaluated.

## 3. Results and Lessons Learned

*a. Model performance and forecasts of opportunity*

Figure 3 shows the skill of the Whole Coast (green), Cluster (orange), and Point (blue) tasks' mean CRPS as a function of location along the coast (Figure 3a) and the distributions of CRPS for each task aggregated over all samples and all 35 tide gauge gridpoints (Figure 3b). All tasks (Figure 3a, colored dotted lines) outperform climatology (Figure 3a, black dotted line) based on mean CRPS over all test set samples by similar margins. The mean CRPS for the Whole Coast task's confident predictions (green solid line) are comparable with the average CRPS for all test set samples (green dotted line) for a majority of the locations, and thus this prediction task is the least suited to select FOOs. For the Cluster and Point tasks, the confident predictions (orange and blue solid lines) have a much lower CRPS than the entire test set (dotted orange and blue solid lines), particularly north of 40N. In other words, the Cluster task is as skillful as the Point task at identifying FOOs. Figure 3b shows that when all test set points are considered, there are substantial differences in the IQR of the CRPS distributions, with all tasks outperforming climatology (compare black and colored dotted boxes, Figure 3b) and the confident predictions further improving upon CRPS for the Cluster and Point tasks (compare orange and blue dotted and filled colored boxes, Figure 3b).

When climate scientists, especially those dealing with model data or large ensembles, are setting up CNN-based prediction systems, one choice we must make is how many CNNs to use to represent the output data. We know from past experience that training one model for every point of interest can become cumbersome (Brettin et al., 2025; Davenport et al., 2024; Gordon et al., 2023; Mayer & Barnes, 2021; Toms et al., 2021), and this work was no exception. For the Cluster task, we trained 15 CNNs totaling $1.8*10^7$ tunable parameters to represent the NEPC, while representing the entire coast with the Point prediction task would have taken over a billion tunable parameters ($1.7*10^6$ tunable parameters for each of 650 CNNs).

Thus, in this work we learned an important lesson. Clustering is a skillful and efficient approach that leverages temporal and spatial covariance to preprocess the data for Week 3 prediction of NEPC sea level.

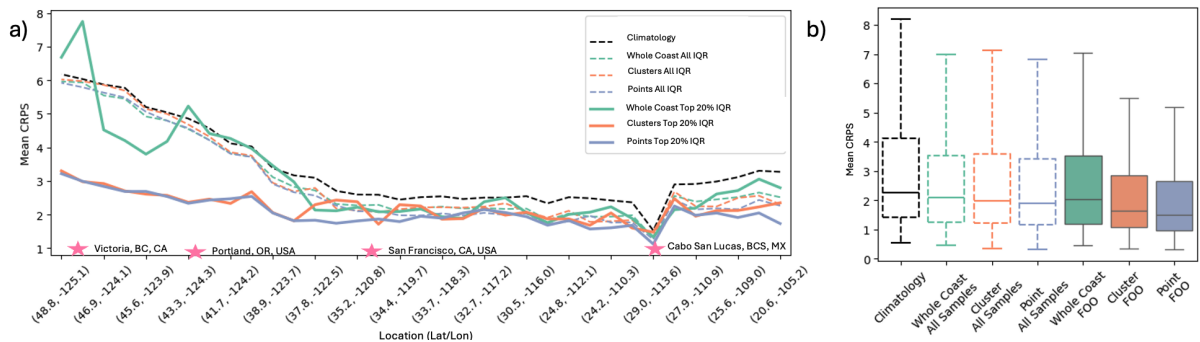

Figure 3. Model CRPS. a) shows model CRPS as a function of location along the NEPC, with the southernmost end of the domain on the right, and the northernmost end of the domain on the left. The coastline has been flattened, so pink stars indicate the approximate locations of major cities to orient the reader. Dotted lines indicate mean CRPS over the test set for the climatology (black dotted), whole coast (green dotted), cluster (orange dotted) and point (blue

dotted), and solid lines indicate mean CRPS over the subset of the test set with the 20% narrowest IQR for the whole coast (green solid), whole coast (orange solid) and point (blue solid) models. b) shows the distribution of CRPS scores for the (subsets of) the test sets, colors and line styles as above. N=2914 for the whole test set, N=583 for the top 20%.

## 4. Discussion

In this work, we explore three ways to partition the NEPC (between 20N and 50N) for predicting sea level anomalies at Week 3 using a simple CNN. We found that pre-processing the coastal points using clustering (Cluster task, Falasca et al., 2024) leads to CNNs which predict sea level with comparable skill to CNNs which were only trained to predict one gridpoint at a time (Point task). This is consistent with work which has shown that clustering locations via k-means can improve one month to three year predictions of sea level in the Mediterranean Sea (Radin et al., 2024) and with work which has shown that multidecadal sea level trend predictions can be improved by spectral clustering (Sinha et al., 2024). Additionally, Oelsmann et al., (2024) and Radin et al., (2024) showed that there are regions of coherent variability on monthly timescales (coastlines such as the Western North American Coast, patches of the Mediterranean), and this work indicates that within the NEPC sub-monthly coherent variability can be leveraged to improve predictability and identify forecasts of opportunity. Our lessons learned underscore the importance of pre-clustering as a tool for dimensionality reduction in geophysical systems; we can and should leverage regions of coherent variability to improve predictability.

*Acknowledgements*

This work was supported by NOAA grant NA24OARX431C002 and NOAA-OAR-CPO-2023-2007559. The authors would also like to thank Dillon J. Amaya for his assistance with processing CESM-HR PI-Control output.

*Data Availability Statement*

Code to generate figures and perform analysis can be found here: https://github.com/lthapa42/lthapa, and raw CESM-HR data may be found here: https://gdex.ucar.edu/datasets/d651029/#